\documentclass[default,iicol,sn-aps]{sn-jnl}

\usepackage{graphicx}%
\usepackage{multirow}%
\usepackage{amsmath,amssymb,amsfonts}%
\usepackage{amsthm}%
\usepackage{mathrsfs}%
\usepackage[title]{appendix}%
\usepackage{xcolor}%
\usepackage{textcomp}%
\usepackage{manyfoot}%
\usepackage{booktabs}%
\usepackage{algorithm}%
\usepackage{algorithmicx}%
\usepackage{algpseudocode}%
\usepackage{listings}%
\usepackage{siunitx}

\usepackage[caption=false]{subfig}

\usepackage{todonotes}

\usepackage{xcolor} 
\definecolor{azure}{rgb}{0.0, 0.5, 1.0}
\definecolor{darkgreen}{rgb}{0.0, 0.5, 0.0}
\definecolor{amaranth}{rgb}{0.9, 0.17, 0.31}
\definecolor{cadetgrey}{rgb}{0.57, 0.64, 0.69}
\definecolor{aureolin}{rgb}{0.99, 0.93, 0.0}

\usepackage{pgfplots}
\pgfplotsset{compat=newest}

\newcommand{\cpp}[1]{\lstinline[language=c++,keywordstyle ={\color{azure}},morekeywords={co_return,co_yield,co_await,ensure_started,sync_wait,start_detached,any_sender}]{#1}}

\newcommand{\header}{\noindent
\rule{\textwidth}{0.4pt}}

\theoremstyle{thmstyleone}%
\theoremstyle{thmstyletwo}%

\theoremstyle{thmstylethree}%

\begin{document}

\title[Shared memory parallelism in Modern C\texttt{++} and HPX]{Shared memory parallelism in Modern C\texttt{++} and HPX}


\author*[1,2]{\fnm{Patrick} \sur{Diehl}}\email{patrickdiehl@lsu.edu}

\author[1]{\fnm{Steven R.} \sur{ Brandt}}\email{sbrandt@cct.lsu.edu}

\author[1]{\fnm{Hartmut} \sur{Kaiser}}\email{hkaiser@cct.lsu.edu}

\affil*[1]{\orgdiv{Center of Computation \& Technology}, \orgname{Louisiana State University}, \orgaddress{\street{Digital Media Center}, \city{Baton Rouge}, \postcode{70803}, \state{LA}, \country{USA}}}

\affil[2]{\orgdiv{Department of Physics and Astronomy}, \orgname{Louisiana State University}, \orgaddress{\street{Street}, \city{Baton Rouge}, \postcode{70803}, \state{LA}, \country{USA}}}


\abstract{Parallel programming remains a daunting challenge, from struggling to express a parallel algorithm without cluttering the underlying synchronous logic to describing which devices to employ to calculate correctness. Over the years, numerous solutions have arisen, requiring new programming languages, extensions to programming languages, or adding pragmas. Support for these various tools and extensions is available to varying degrees. In recent years, the C\texttt{++} standards committee has worked to refine the language features and libraries needed to support parallel programming on a single computational node. Eventually, all major vendors and compilers will provide robust and performant implementations of these standards. Until then, the HPX library and runtime provide cutting-edge implementations of the standards and proposed standards and extensions. Because of these advances, it is now possible to write high performance parallel code without custom extensions to C\texttt{++}. We provide an overview of modern parallel programming in C\texttt{++}, describing the language and library features and providing brief examples of how to use them.}

\keywords{C\texttt{++}, HPX, AMT, Parallelism}



\maketitle

\section{Introduction}
Parallel programming is essential to modern software development and is supported in recent programming languages like Julia or Rust. However, in older languages such as C\texttt{++}, parallel programming features were not originally included as language or library features.

To address this omission, POSIX threads~\cite{butenhof1997programming}, so-called \textit{pthreads}, a C library, was created for the Unix operating system. The application program interface (API) for \textit{pthreads} was defined by the \textit{POSIX.1C} thread extension (\textit{IEEE Std 1003.1c-1995}). Likewise, with the C\texttt{++} \num{11} standard~\cite{cxx11_standard}, \cpp{std::thread} was added in C\texttt{++} as a low level interface. At a higher abstraction layer, \cpp{std::async} and \cpp{std::future} for asynchronous programming were added.

In addition, the standard supplied parallel programming utilities, which aided in writing parallel programs (\emph{e.g.}\ smart pointers and lambda functions). With the C\texttt{++} \num{14} standard~\cite{cxx14_standard}, these utilities were further augmented with generic lambda functions and shared mutexes.

To make parallel programming more accessible and less error-prone, the C\texttt{++} \num{17} standard~\cite{cxx17_standard} introduced parallel algorithms, allowing programmers to execute most of the algorithms from the C\texttt{++} 98 standard in parallel (\emph{e.g.}\ \cpp{std::sort} or \cpp{std::reduce}) 

Coroutines were added with the C\texttt{++} \num{23} standard to support asynchronous programming. The keywords \cpp{co_return}, \cpp{co_yield}, and \cpp{co_await} added functionality to suspend and resume functions. Also, in the C\texttt{++} \num{23} standard, the \textit{ranges} library was added, which can be seen as the generalization and extension of the algorithm
library. Finally, utilities such as semaphores, latches, and barriers were added.
Soon, it is expected that \cpp{std::async} will become deprecated to be succeeded by the sender and receiver library (which has yet to be accepted).

\begin{figure*}[tbp]
    \centering
 \begin{tikzpicture}

\draw[->,thick] (0,0) -- (11,0);
\node[above] at (1,0.125) {C\texttt{++} \num{11}};
\node[above] at (4,0.125) {C\texttt{++} \num{14}};
\node[above] at (7,0.125) {C\texttt{++} \num{17}};
\node[above] at (10,0.125) {C\texttt{++} 20};

\node[below] at (1,-0.125) {\cpp{std::thread}};
\node[below] at (1,-0.125-0.5) {std::async};
\node[below] at (1,-0.125-1) {Smart pointer};
\node[below] at (1,-0.125-1.5) {Lambda functions};
\node[below] at (4,-0.125) {Generic lambda};
\node[below] at (4,-0.125-0.5) {shared mutex};
\node[below] at (7,-0.125) {Parallel };
\node[below] at (7,-0.125-0.5) {algorithms};
\node[below] at (10,-0.125) {Coroutines};
\node[below] at (10,-0.125-0.5) {Ranges};
\node[below] at (10,-0.125-1.0) {Semaphores};
\node[below] at (10,-0.125-1.5) {Latch};
\node[below] at (10,-0.125-2) {Barrier};

\end{tikzpicture}
    \caption{timeline of the parallel features added to the C\texttt{++} standard from C\texttt{++} \num{11} to C\texttt{++} 20.}
    \label{fig:cpp:overview}
\end{figure*}

The C\texttt{++} standard library for parallelism and concurrency (HPX) implements all the latest features, both proposed and accepted in the C\texttt{++} standard. In addition, HPX provides extensions to the functionality of the standard, providing mechanisms for distributed parallel programming, alternative ways to create asynchrony, and more.

What is HPX? HPX is an asynchronous many-task runtime system. HPX employs light-weight (user-level) threads that are cooperatively scheduled on top of operating system threads and performs context switches to enable blocked threads to get back to work.

For more details about HPX, we refer to Section~\ref{sec:hpx}. Because HPX conforms to the C\texttt{++} standard, any conforming C\texttt{++} code can be easily converted to HPX by changing some headers and namespaces. To conclude, while single node parallelism is included in the C\texttt{++} standard and no external libraries or language extensions are needed, HPX provides a reliable way to stay on the cutting edge of the standard.

In this paper, we will introduce asynchronous programming, parallel algorithms, and coroutines, senders and receivers (see P2300), and compare the performance between (standard) C\texttt{++} using operating system threads and HPX using light-weight threads. Finally, we will discuss the benefits of each approach.

The paper is structured as follows: Section~\ref{sec:related:work} gives a brief overview of related work. Section~\ref{sec:hpx} introduces HPX and the features described in this paper. In Section~\ref{sec:approaches}, four approaches to implementing the Taylor series of the natural logarithm are provided. Section~\ref{sec:comparison} compares the programming paradigms used in these approaches. Section~\ref{sec:performance} compares the performance of the approaches on Intel, AMD, and A64 FX CPUs. Finally, Section~\ref{sec:conclusion} concludes the work.

\section{Related work}
\label{sec:related:work}
In the past, parallelism in C\texttt{++} was usually achieved by using the OpenMP~\cite{chandra2001parallel} and Cilk~\cite{Leiserson2011} as language extensions. Alternatively, Intel Thread Building Blocks (TBB), Microsoft Parallel Patterns Library (PPL) provided access to parallelism through libraries. More recently, Kokkos~\cite{CarterEdwards20143202} has provided a library interface for parallel and heterogeneous computing. While all these approaches have different advantages, they also have different interfaces, and none are part of the C\texttt{++} standard. Conforming to the standard might mean that future versions of a conforming code compile and run more reliably, and this is a critical consideration among many in constructing a new parallel program or adding parallelism to an existing code.

Another longtime player in the asynchronous many-thread library arena is Charm\texttt{++}~\cite{kale1993charm++}. Like HPX, Charm++ also provides facilities for distributed programming (for which, at present, the C\texttt{++} provides no standard). For a comparison of Charm\texttt{++} and HPX with OpenMP and MPI (a widely accepted standard for distributed parallel programming) using Task\ Bench, we refer to~\cite{wu2023quantifying}. Other notable AMTS are: Chapel~\cite{chamberlain2007parallel}, X10~\cite{ebcioglu2004x10}, and UPC\texttt{++}~\cite{zheng2014upc++}. For a more detailed comparison of AMTs, we refer to~\cite{thoman2018taxonomy}. Table~\ref{tab:support:approaches} lists the support of approaches, namely, futures and futurization (Section~\ref{sec:approaches:future}), coroutines (Section~\ref{sec:approaches:coroutines}), parallel algorithms (Section~\ref{sec:approaches:parallel}), and senders \& receivers (Section~\ref{sec:approaches:sender:receiver}); by other AMTs. We left X10 out, since the last release was made in 2019.
Charm\texttt{++} provides futures but not coroutines. 
The functionality similar to senders and receivers is available, however. A \emph{Chare} can be used somewhat like a scheduler, and a Charm\texttt{++} callback can provide similar functionality to \texttt{then()}.

Chapel provides futures. Parallel algorithms are partially supported, \emph{e.g.}\ parallel \cpp{for} loops. Coroutines and sender \& receivers are not supported. UPC\texttt{++} has futures but does not support the other features. 

\begin{table*}[t]
    \centering
    \caption{Availability of the studied approaches in other run time systems. However, solely HPX is studied in this paper. Therefore, HPX is the basis for the comparison of the features. With \texttildelow~ we indicated that the features are only partially supported. }
    \begin{tabular}{l|cccc} \toprule
     Approach    & Futurization & Coroutines & Parallel Algorithms & Sender \& Receivers  \\ \midrule
     \textcolor{azure}{HPX} &  \textcolor{azure}{\checkmark} &  \textcolor{azure}{\checkmark} &  \textcolor{azure}{\checkmark} &  \textcolor{azure}{\checkmark}\\
     Charm\texttt{++} & \checkmark & \texttildelow & \text{\sffamily X} & \text{\sffamily X} \\
     Chapel & \checkmark & \text{\sffamily X}  & \texttildelow & \text{\sffamily X}  \\
     UPC\texttt{++} & \checkmark & \text{\sffamily X} & \text{\sffamily X} & \text{\sffamily X}  \\ \bottomrule
    \end{tabular}
    \label{tab:support:approaches}
\end{table*}

\section{HPX}
\label{sec:hpx}

HPX~\cite{Kaiser2020} is an Asynchronous Many-task Runtime System (AMT) that exposes an ISO C\texttt{++} standards conforming API for shared memory parallel programming, and extensions to that API library that enable distributed computing. 
This API enables asynchronous parallel programming through futures, senders and receivers, channels, and other synchronization primitives. 
This API also eases the burden on a new programmer while learning how to use HPX. 
It also guarantees application portability in terms of code and performance.
HPX employs a user-level threading system that can fully exploit available parallel resources through fine-grain parallelism on various contemporary and emerging high-performance computing architectures. 
HPX makes it possible to create scalable parallel applications that expose excellent parallel efficiency and high resource utilization. 
HPX's asynchronous programming model enables intrinsic overlapping of computation and communication, prefers moving work to data over moving data to work, and does so while exposing minimal overheads.

In the context of this paper, we focus on assessing the performance of HPX's implementation of futures and parallel algorithms as mandated by the C\texttt{++} \num{17}, 20, and \num{23} standards.

\section{Approaches}
\label{sec:approaches}
To showcase the various approaches to shared memory parallelism, we will implement the Taylor series for the natural logarithm in parallel. The Maclaurin series for the natural logarithm $ln$ with the basis $e$ reads as
\begin{align}
    & ln(1+x) = \notag\\
    &\sum\limits_{n=1}^\infty (-1)^{n+1} \frac{x^n}{n} = x - \frac{x^2}{2} + \frac{x^3}{3} - \ldots, \text{with } \vert x \vert < 1 \text{.}
    \label{eq:taylor:series}
\end{align}
For simplicity, we will omit the main method and all headers from the code examples. However, we will mention the specific headers in the text, and we provide the complete code for all examples on GitHub\textsuperscript{\textregistered}.

\subsection{Futures and Futurization}
\label{sec:approaches:future}
The current abstractions for parallel programming in C\texttt{++} are low-level threads \cpp{std::thread}, \cpp{std::async}, and \cpp{std::future}. However, in a future C\texttt{++} standard, it is expected that some of these facilities will become deprecated and will be replaced by sender and receivers. HPX, however, will continue to support an extended version of futures which share many of the capabilities of senders and receivers, including a \texttt{then()} method, a \texttt{when\_all()} method, executors, and so on.

Futures represent a proxy for a result that may not yet be computed and provide a relatively intuitive way to express asynchronous computations. The C\texttt{++} standard allows programmers to retrieve the value of futures using the \cpp{get()} method, but HPX allows programmers to attach a continuation to the future using the \cpp{then(std::function<T>)} method. This capability, combined with a \cpp{when_all()} method for waiting for future groups, makes it possible to write asynchronous subroutines and algorithms that never block. This is an essential consideration for libraries that rely on a pool of workers to carry out parallel computations. Blocking one or more of them might lead not only to slower code, but also blocked code. Routines that are rewritten in this way to run in parallel but without calling \cpp{get()} are said to be {\em futurized}. As of this writing, futurized code is only possible with HPX, and not with the C\texttt{++} standard.

Listing~\ref{algorithm:taylor:future:hpx} shows the implementation. The amount of work is divided equally among threads. In Line~\ref{algorithm:taylor:future:hpx:launch}, a lambda function is launched to act on each chunk of work asynchronously and an \cpp{hpx::future<double>} is returned. Note that we do not need to wait for the lambda function to be finished, and the \cpp{for} loop proceeds. This happens because the \cpp{hpx::future} is a placeholder for the result of the lambda function, freeing us from the need to wait for it to be computed. In Line~\ref{algorithm:taylor:future:hpx:when:all} a barrier is introduced to collect the partial results using \cpp{hpx::when_all}. Here, the HPX runtime waits until all futures are ready, which means that the computation in the lambda function has finished. In Line~\ref{algorithm:taylor:future:hpx:then} we specify which lambda function is called. We use the \cpp{.get()} function to collect all the partial results. If the result is not ready, HPX would wait here for the result to be ready. However, due to the \cpp{hpx::when_all} all results are ready. In Line~\ref{algorithm:taylor:future:hpx:get2}, we need to call \cpp{.get()} since \cpp{hpx::when_all} returns a future for integration in the asynchronous dependency graph.

\begin{lstlisting}[language=c++,caption={Parallel implementation of the natural logarithm using \cpp{hpx::async} and \cpp{hpx::future}.\label{algorithm:taylor:future:hpx}},escapechar=|,basicstyle=\small\ttfamily,float=*tbp] 
double run(size_t n, size_t num_threads, double x) {
  std::vector<double> parts(n);
  std::iota(parts.begin(), parts.end(), 1);

  size_t partition_size = n / num_threads;

  std::vector<hpx::future<double>> futures;
  for (size_t i = 0; i < num_threads; i++) {
    size_t begin = i * partition_size;
    size_t end = (i + 1) * partition_size;
    if (i == num_threads - 1) end = n;

    hpx::future<double> f = hpx::async(
        [begin, end, x, &parts]() -> double { |\label{algorithm:taylor:future:hpx:launch}|
      std::for_each(parts.begin() + begin,
                    parts.begin() + end, [x](double& e) {
        e = std::pow(-1.0, e + 1) * std::pow(x, e) / (e);
      });

      return hpx::reduce(parts.begin() + begin,
                         parts.begin() + end, 0.);
    });

    futures.push_back(std::move(f));
  }

  double result = 0;

  hpx::when_all(futures) |\label{algorithm:taylor:future:hpx:when:all}|
      .then([&](auto&& f) { | \label{algorithm:taylor:future:hpx:then}|
        auto futures = f.get(); | \label{algorithm:taylor:future:hpx:get}|

        for (size_t i = 0; i < futures.size(); i++)
            result += futures[i].get();
      })
      .get();   | \label{algorithm:taylor:future:hpx:get2}|

  return result;
}
\end{lstlisting}

\subsection{Coroutines}
\label{sec:approaches:coroutines}
With C\texttt{++} 20 coroutines, functions that can be suspended and resumed were added. The three following \cpp{return} types are available for coroutines: \cpp{co_return} which is similar to \cpp{return}, but the function is suspended; \cpp{co_yield} returns the expression to the caller and suspends the current coroutine; and \cpp{co_await} which suspends the coroutine and returns the control to the caller.

A coroutine version of Listing~\ref{algorithm:taylor:future:hpx} can be found in Listing~\ref{algorithm:taylor:future:hpx:coroutine} In Line~\ref{algorithm:taylor:future:hpx:coroutine:run} of Listing~\ref{algorithm:taylor:future:hpx:coroutine} we define the function \cpp{run} as our coroutine by having it return an \texttt{hpx::future}. Next, we copied the code from Listing~\ref{algorithm:taylor:future:hpx} for the evaluation of the Taylor series, however, we changed three lines to use the new coroutine features. First, in Line~\ref{algorithm:taylor:future:hpx:coroutine:coawait:1}, we use \cpp{co_await} while we wait for all futures. Second, in Line~\ref{algorithm:taylor:future:hpx:coroutine:coawait:2}, we use \cpp{co_await} to collect the partial results of all futures. Note in Listing~\ref{algorithm:taylor:future:hpx}, we had to call \cpp{.get()} here to wait for the futures. Third, in Line~\ref{algorithm:taylor:future:hpx:coroutine:coawait:2}, we call \cpp{co_return} at the end of our coroutine. Note that internally HPX will call \cpp{.get()} where we use \cpp{co_await}, so the code is easier to read but will not run faster.

\begin{lstlisting}[language=c++,caption={Example for the computation of the Taylor series for the natural logarithm using HPX's futures and coroutines.\label{algorithm:taylor:future:hpx:coroutine}},escapechar=|,keywordstyle ={\color{azure}},morekeywords={co_return,co_yield,co_await,HPX_PLAIN_ACTION},basicstyle=\small\ttfamily,float=*]
#include <coroutine>
|\header|
hpx::future<double> run(size_t n,
                        size_t num_threads,
                        double x) { |\label{algorithm:taylor:future:hpx:coroutine:run}|
  std::vector<double> parts(n);
  std::iota(parts.begin(), parts.end(), 1);

  size_t partition_size = n / num_threads;

  std::vector<hpx::future<double>> futures;
  for (size_t i = 0; i < num_threads; i++) {
    size_t begin = i * partition_size;
    size_t end = (i + 1) * partition_size;
    if (i == num_threads - 1) end = n;

    hpx::future<double> f = hpx::async(
        [begin, end, x, &parts]() -> double {
      std::for_each(parts.begin() + begin,
                    parts.begin() + end, [x](double& e) {
        e = std::pow(-1.0, e + 1) * std::pow(x, e) / (e);
      });

      return hpx::reduce(parts.begin() + begin,
                         parts.begin() + end, 0.);
    });

    futures.push_back(std::move(f));
  }

  double result = 0;

  auto futures2 = co_await hpx::when_all(futures); |\label{algorithm:taylor:future:hpx:coroutine:coawait:1}|

  for (size_t i = 0; i < futures2.size(); i++)
    result += co_await futures2[i];  |\label{algorithm:taylor:future:hpx:coroutine:coawait:2}|

  co_return result; |\label{algorithm:taylor:future:hpx:coroutine:coreturn}|
}
\end{lstlisting}

\subsection{Parallel algorithms}
\label{sec:approaches:parallel}
The algorithms within the C\texttt{++} standard library introduced with the C\texttt{++} 98 standard were extended with parallel execution in the C\texttt{++} \num{17} standard. Listing~\ref{algorithm:taylor:parallel:alg} shows the complete code. In Line~\ref{algorithm:taylor:parallel:alg:for} we use the algorithm \cpp{std::for_each} to iterate over each element of the \cpp{std::vector} to evaluate the value $x$ of the Taylor series. In Line~\ref{algorithm:taylor:parallel:alg:sum} the algorithm \cpp{std::reduce} is used to compute the sum of all evaluations. Note that the only difference between the parallel version and the original  C\texttt{++} 98 standard is the first argument of both algorithms, the execution policy. The following execution policies in the header \cpp{#include <execution>}~\cite{p2300} are currently available:
\begin{itemize}
    \item \cpp{std::execution::par}: The algorithm is executed in parallel using multiple operating system threads.
    \item \cpp{std::execution::seq}: The algorithm is executed in parallel using one operating system thread.
    \item \cpp{std::execution::par_unseq}: The algorithm is executed in parallel using multiple operating system threads and vectorization for additional optimizations.
\end{itemize}
Note that this is still an experimental feature and, as of this writing, only the GNU compiler collection (GCC) $\geq 9$ and Microsoft Visual C\texttt{++} compiler $\geq 15.7$ support this feature. Intel's One API compiler uses Thread Building Blocks (TBB) to implement this feature.

\begin{lstlisting}[language=c++,caption={Implementation of the Taylor series of the natural logarithm using C\texttt{++} parallel algorithms.\label{algorithm:taylor:parallel:alg}},escapechar=|,basicstyle=\small\ttfamily,float=*] 
#include <iostream>
#include <future>
#include <vector>
#include <algorithm>
#include <numeric>
#include <execution>
#include <cmath>
|\header|
double run(size_t n, size_t num_threads, double x) {
    std::vector<double> parts(n); 
    std::iota(parts.begin(), parts.end(), 1); 

    std::for_each(std::execution::par,
                  parts.begin(),
                  parts.end(), [x](double& e) { |\label{algorithm:taylor:parallel:alg:for}|
        e = std::pow(-1.0, e+1) * std::pow(x, e) / e;
    });

    double result = std::reduce(std::execution::par,
                                parts.begin(),
                                parts.end(), 0.); |\label{algorithm:taylor:parallel:alg:sum}|
    return result;
}
\end{lstlisting}

The same functionality for \texttt{execution} of parallel algorithms is available within HPX. 

\subsubsection{Additional HPX features}
\label{sec:approaches:parallel:hpx:features}
However, HPX extends the current features available in the C\texttt{++} \num{17} standard, allowing execution policies with chunk sizes to specify the amount of work each thread is operating on at once. The following chunk sizes are available:
\begin{itemize}
    \item \cpp{hpx::execution::static_chunk_size}: The container elements are divided into pieces of a given size and then assigned to the threads.
    \item \cpp{hpx::execution::auto_chunk_size}: Chunk size is determined after 1\% of the total container elements were executed. 
    \item \cpp{hpx::execution::dynamic_chunk_size}: Dynamically scheduled among the threads and if one thread is done it gets dynamically assigned a new chunk.
\end{itemize}

For details about the effect of chunk sizes on performance, we refer to~\cite{grubel2015performance}. A machine learning approach to determining chunk size is presented here~\cite{shirzad2019scheduling,khatami2017hpx}. With respect to vectorization, HPX provides the execution policy \cpp{hpx::execution::simd} to execute the algorithm using vectorization. In addition, HPX provides a combined execution policy \cpp{hpx::execution::par_simd} to combine parallelism and vectorization. Here, \cpp{std::experimental:simd}~\cite{9651210}, Vc~\cite{kretz2012vc}, and Eve are possible backends. Furthermore, HPX's parallel algorithms can be combined with asynchronous programming. Here, an \cpp{hpx::future} is returned and can be integrated into HPX's asynchronous execution graph.

Listing~\ref{algorithm:taylor:parallel:alg:hpx} shows the usage of the chunk size feature. In Line~\ref{algorithm:taylor:parallel:alg:hpx:cunk:size} a static chunk size of ten is defined and passed to the \cpp{hpx::for_each} in Line~\ref{algorithm:taylor:parallel:alg:hpx:cunk} by using \cpp{.with()}. In Line~\ref{algorithm:taylor:parallel:alg:hpx:future} the parallel algorithm \cpp{hpx::reduce} is wrapped into a future, which can be integrated within HPX's asynchronous dependency graph. 

\begin{lstlisting}[language=c++,caption={Implementation of the Taylor series of the natural logarithm using parallel algorithms.\label{algorithm:taylor:parallel:alg:hpx}},escapechar=|,basicstyle=\small\ttfamily,float=*tpb] 
#include <hpx/execution/executors/static_chunk_size.hpp>
|\header|
double run(size_t n, size_t num_threads, double x) {
    hpx::execution::static_chunk_size scs(10); |\label{algorithm:taylor:parallel:alg:hpx:cunk:size}|
    std::vector<double> parts(n); 
    std::iota(parts.begin(), parts.end(), 1); 
    hpx::for_each(
         hpx::execution::par.with(scs),
         parts.begin(), parts.end(), |\label{algorithm:taylor:parallel:alg:hpx:cunk}|
         [x](double& e) { e = std::pow(-1.0, e+1) * std::pow(x, e) / e; });

    hpx::future<double> f = |\label{algorithm:taylor:parallel:alg:hpx:future}|
         hpx::reduce(hpx::execution::par(hpx::execution::task),
                     parts.begin(), parts.end(), 0.);
    return f.get();
}

int main() {
    int n = 1000;
    double x = .1;
    double result = run(n,10,x);
    std::cout << "Result is: " << result << std::endl;
    std::cout << "Difference of Taylor and C\texttt{++} result "
              << result - std::log1p(x) << " after "
              << n << " iterations." << std::endl;
}
\end{lstlisting}

\subsection{Senders and Receivers}
\label{sec:approaches:sender:receiver}
A new framework for writing parallel codes is currently being debated by the C\texttt{++} standards committee: senders and receivers. One of the goals of this framework is to make it easier to execute codes on heterogeneous devices. The various devices are expressed as {\em schedulers}. In principle, these could be GPUs, different NUMA domains, or arbitrary groups of cores.

Each step of a calculation is expressed as a {\em sender}. Senders are typically chained together using the pipe operator in analogy to the bash shell. Values, error conditions (exceptions), as well as requests to stop a computation, can be carried through the pipeline.

By default, building the pipeline does nothing. Execution begins only when \cpp{ensure_started()}, \cpp{sync_wait()}, or \cpp{start_detached()} is called.

Receivers are usually implicit, hidden in the call to \cpp{sync_wait()} at the end.

We note that this proposal was not accepted into the C\texttt{++} \num{23} standard, partly because it was proposed too close to the deadline. It may also need further development.
In our experiments writing short codes to use senders and receivers, we attempted to write a recursive Fibonacci routine that took a sender as input and produced a sender as output and did not itself call \cpp{sync_wait()} to get the result. In order to write it, we needed to make use of the \cpp{any_sender<T>} class provided in the HPX implementation but not specified in the standard yet. Whether additions of this kind turn out to be necessary, or whether the proposal itself will ultimately be accepted, remains for the committee to decide.

\begin{lstlisting}[language=c++,caption={Implementation of the Taylor series of the natural logarithm using sender and receivers.\label{algorithm:taylor:sender:receiver:alg:hpx}},escapechar=~,keywordstyle={\color{azure}},morekeywords={co_return,co_yield,co_await,ensure_started,sync_wait,start_detached,any_sender},basicstyle=\small\ttfamily,float=*tbp] 
#include <hpx/execution.hpp>

using namespace hpx::execution::experimental;

template <typename T> concept sender = is_sender_v<T>;

namespace tt = hpx::this_thread::experimental;
~\header~
double run(size_t n, size_t num_threads, double x) {
  thread_pool_scheduler sch{};

  size_t partition_size = n/num_threads;
  std::vector<double> partial_results(partition_size);

  sender auto s = schedule(sch) |
    bulk(num_threads, [&](auto i) {
      size_t begin = i * partition_size;
      size_t end = (i + 1) * partition_size;
      if (i == num_threads - 1) end = n;
      double partial_sum = 0;
      for(int i=begin; i <= end; i++) {
        double e = i+1;
        double term = std::pow(-1.0, e+1) * std::pow(x, e) / e;
        partial_sum += term;
      }
      partial_results[i] = partial_sum;
    }) | 
    then([&]() {
      double sum = 0;
      for(int i=0;i<partition_size;i++)
        sum += partial_results[i];
      return sum;
    });
  auto [result] = *tt::sync_wait(std::move(s));
  return result;
}

int main() {
  double x = .1;
  double r = run(10000,10,x);
  double a = log(1+x);
  std::cout << "r=" << r << " ~ " << a
            << " => " << fabs(r-a) << std::endl;
}

\end{lstlisting}

\section{Comparison of the approaches}
\label{sec:comparison}

In the previous section, the focus was on how to implement the Taylor series for the natural logarithm, see Equation~\eqref{eq:taylor:series}, using the various approaches.

The fundamental difference in the approaches lies in where the various codes block and how much overhead they introduce. For the standard library, calls to \texttt{future.get()} will potentially block. In our parallel future listing~\ref{algorithm:taylor:future:hpx} we use \texttt{when\_all()} which defers most of the calls to \texttt{get()} until all futures are ready. Thus, ony the final call to \texttt{get()} can block.

For HPX, anything that would normally block will instead be suspended and switched out, similar to what C\texttt{++} Coroutines would do.

Which leads us to the explicit coroutine code. Performing the suspend and resume operations are sure to introduce overheads, but they should not be as large as they seem to be from our data. This was easily the slowest version of the code. See Listing~\ref{algorithm:taylor:future:hpx:coroutine}.

The parallel library approach does not attempt to suspend or resume, it performs a simple fork-join on evenly divided threads. This avoids the overheads of suspending and resuming, but potentially causes threads to wait unnecessarily at the joins. Listing~\ref{algorithm:taylor:parallel:alg} shows this approach.

Finally, senders and receivers, Listing~\ref{algorithm:taylor:sender:receiver:alg:hpx} shows the most recent proposed method of implementing asynchrony in C++. This represents an effort to provide ways to express asynchrony while avoiding the overheads of futures and coroutines. Our data shows that it is fairly successful as, for most core counts, this was the fastest.



\begin{figure}[tb]
    \centering
    \begin{tikzpicture}
    \node[] at (3,1.5) {Concepts} ;
    \draw[rounded corners,cadetgrey] (0,0) rectangle (2,1) node[pos=.5] {\textcolor{black}{Concurrency}};
    \draw[rounded corners,cadetgrey] (4,0) rectangle (6,1) node[pos=.5] {\textcolor{black}{Parallelism}};
    \draw[<->,cadetgrey] (2,0.5) -- (4,0.5);
    \draw[dashed] (0,-0.5) -- (7,-0.5);
     \draw[cadetgrey] (0,-1) rectangle (2,-2) node[pos=.5, align=center] {\textcolor{black}{Future} \\ \textcolor{black}{Async}};
     \draw[cadetgrey] (5,-1) rectangle (7,-2) node[pos=.5, align=center] {\textcolor{black}{Parallel} \\ \textcolor{black}{Algorithms}};
     \draw[cadetgrey] (2.5,-1) rectangle (4.5,-2.25) node[pos=.5, align=center] {\textcolor{black}{Senders} \\ \textcolor{black}{\&} \\ \textcolor{black}{Receivers}  };
    \draw[->] (1,0) -- (1,-1);
    \draw[->] (1,0) -- (3.5,-1);
    \draw[->] (5,0) -- (6,-1);
    \draw[->] (5,0) -- (3.5,-1);
    \node[] at (3,-3) {C\texttt{++} features} ;
    \end{tikzpicture}
    \caption{On the top: The two concepts, namely, concurrency and parallelism. Where concurrency describes the structure of the sequences, like suspending tasks ad resuming tasks, and parallelism describes the execution. On the bottom, the approaches to implement concurrency and  parallelism in Modern C\texttt{++}: Futures + Async (Section~\ref{sec:approaches:future}); Parallel Algorithms (Section~\ref{sec:approaches:parallel}); Senders and Receivers (Section~\ref{sec:approaches:sender:receiver}); and Coroutines + Async (Section~\ref{sec:approaches:coroutines}).}
    \label{fig:parallelism:concurrency}
\end{figure}

Let us transfer this example to the C\texttt{++} programming language. Figure~\ref{fig:parallelism:concurrency} shows the classification of the C\texttt{++} approaches concerning parallelism and concurrency. For parallelism, the C\texttt{++} standard provides three approaches. First, the parallel algorithms introduced with the C\texttt{++} 17 standard, see Section~\ref{sec:approaches:parallel}. However, the parallel algorithms are very restricted since these algorithms operate on the elements of containers, \emph{e.g.} \cpp{std::vector}. Some algorithms like \cpp{std::sort} or \cpp{std::find_if} are customizable by providing compare operators like \cpp{std::greater<double>()}  or providing functions or lambda functions.

To summarize, the algorithms (or parallel algorithms) are good for operating on containers in sequential or parallel using execution policies. A more flexible option is asynchronous programming using futures. The interface \cpp{std::async} and \cpp{std::future} and their counterparts \cpp{hpx::async} and \cpp{hpx::future} are abstraction interfaces for low level programming using \cpp{std::threads} and \cpp{hpx::thread}, respectively. The principle for futurization is that the work is split into partitions and each thread works on its assigned partition. Here, the programmer distributes the work as partitions to the threads. Furthermore, HPX allows combining (parallel) algorithms and asynchronous programming by asynchronously launching the algorithms while returning a future. The API for \cpp{std::async} and \cpp{std::future} was introduced with the C\texttt{++} 11 standard, but might be deprecated soon and be replaced with its successor senders and receivers. The current outline is to accept senders and receivers for the C\texttt{++} 26 standard. However, HPX implements the latest proposal. See Section~\ref{sec:approaches:sender:receiver}. 

For concurrency, coroutines were added with the C\texttt{++} 20 standard, see Section~\ref{sec:approaches:coroutines}. The \cpp{co_return}, \cpp{co_yield}, and \cpp{co_await} features were added to suspend and resume coroutines. Note that coroutines themselves do not provide parallelism per se and can be used to create a generator on a single core. Senders and receivers, curiously, provide features for concurrency and parallelism.

Figure~\ref{fig:parallelism:concurrency} could be explicitly extended to HPX. In that case, the parallel algorithms provide (though they are parallel) support concurrency because HPX allows them to return a \cpp{hpx::future}. Furthermore, HPX's parallel algorithms can be integrated within senders and receivers. However, these features are not specified in the C\texttt{++} standard ``yet''.  For comparing parallelism and concurrency in Chapel, Charm\texttt{++}, C\texttt{++}, HPX, Go, Julia, Python, Rust, Swift, and Java for a 1D heat equation solver, we refer to~\cite{diehl2023benchmarking}.

\section{Performance comparison}
\label{sec:performance}
For performance measurements on different CPUs, we compiled all examples using gcc $12.1.0$ for Arm, using gcc $9.2.0$ for AMD and Intel. HPX $1.8.1$ was compiled with the following dependencies: boost $1.78.0$, hwloc $2.2.0$, and jemalloc $5.2.0$. Table~\ref{tab:software} summarizes the versions of dependencies and CPU architectures used for the performance measurements in Figure~\ref{fig:performance}. For all core counts, the code was executed ten times and the median out of these runs is plotted. The error bars show the variances within these ten runs. For some approaches, we observe high variance for HPX on larger node counts.

\begin{table*}[tb]
    \centering
       \caption{Summary of CPU architectures, compilers, and dependencies used for the performance measurements in Figure~\ref{fig:performance}.}
    \begin{tabular}{l|lllll} \toprule
    CPU & gcc & hpx & boost & hwloc & jemalloc \\\midrule
    Intel Xeon Gold 6140     & 9.2.0 & 1.8.1 &  1.78.0 & 2.2.0 & 5.2.0 \\
    AMD EPYC 7543    & 9.2.0 & 1.8.1 &  1.78.0 & 2.2.0 & 5.2.0  \\
    A64FX &  12.1.0 & 1.8.1 &  1.78.0 & 2.2.0 & 5.2.0 \\\bottomrule
    \end{tabular}
    \label{tab:software}
\end{table*}

Figure~\ref{fig:performance} shows the performance obtained for all four of the programming mechanisms presented in this paper: for ARM A64FX, AMD EPYC\textsuperscript{\texttrademark} 7543, and Intel\textsuperscript{\textregistered} Xeon\textsuperscript{\textregistered} Gold 6140, respectively. To create an artificial work load, we computed the Taylor series in Equation~\eqref{eq:taylor:series} for $n=\num{1000000000}$. We used \textit{perf} on the Intel CPU to obtain the floating point operations of \num{100000028581} on a single core. For futures using \cpp{std::future} and \cpp{hpx::future} \protect\subref{fig:performance:arm}, we see that on Arm both implementations perform the same. Similar behavior is obtained for Intel. However, on AMD \cpp{hpx::future} performs better. Here, the overhead of using HPX is negligible. For more details on the overheads of HPX and Charm\texttt{++}, we refer to~\cite{wu2023quantifying}. For HPX's parallel algorithms using \cpp{hpx::for_each} \protect\subref{fig:performance:amd}, AMD performed better as Intel and Arm is around one order of magnitude slower.
The results on Arm64FX are shown in \protect\subref{fig:performance:intel}. The performance of the two more recent C\texttt{++} features is one order of magnitude slower on Arm than on the two other architectures. Senders and receivers showed the best performance on Arm. However, one should not conclude that this paradigm is inherently faster based on this test. Note that we experience some high variation on higher node counts.
More investigation is needed for this feature. For more performance measurements on Rikken's Supercomputer\ Fugaku, we refer to~\cite{diehl2023simulating}.

\begin{figure*}[tbp]
    \centering
      \subfloat[Futurization \label{fig:performance:arm}]{
      \resizebox{0.5\textwidth}{!}
        {
    \begin{tikzpicture}
\begin{axis}[grid,xlabel={\# cores},ylabel={Flop/s},xmax=20,legend pos=north west]
\addplot[black,thick,mark=*, error bars/.cd, y dir=both, y explicit] table [x expr=\thisrowno{0},y expr={100000028581/\thisrowno{2}}, col sep=comma, y error expr={1*(100000028581/\thisrowno{1}-100000028581/\thisrowno{2})}] {ookami_taylor_future_median.csv};
\addplot[gray,thick,mark=*, error bars/.cd, y dir=both, y explicit] table [x expr=\thisrowno{0},y expr={100000028581/\thisrowno{2}}, col sep=comma, y error expr={1*(100000028581/\thisrowno{1}-100000028581/\thisrowno{2})}] {ookami_taylor_future_hpx_median.csv};

\addplot[black,thick,mark=triangle*, error bars/.cd, y dir=both, y explicit] table [x expr=\thisrowno{0},y expr={100000028581/\thisrowno{2}}, col sep=comma, y error expr={1*(100000028581/\thisrowno{1}-100000028581/\thisrowno{2})}] {intel_taylor_future_median.csv};
\addplot[gray,thick,mark=triangle*, error bars/.cd, y dir=both, y explicit] table [x expr=\thisrowno{0},y expr={100000028581/\thisrowno{2}}, col sep=comma, y error expr={1*(100000028581/\thisrowno{1}-100000028581/\thisrowno{2})}] {intel_taylor_future_hpx_median.csv};

\addplot[black,thick,mark=square*, error bars/.cd, y dir=both, y explicit] table [x expr=\thisrowno{0},y expr={100000028581/\thisrowno{2}}, col sep=comma, y error expr={1*(100000028581/\thisrowno{1}-100000028581/\thisrowno{2})}] {amd_taylor_future_median.csv};

\addplot[gray,thick,mark=square*, error bars/.cd, y dir=both, y explicit] table [x expr=\thisrowno{0},y expr={100000028581/\thisrowno{2}}, col sep=comma, y error expr={1*(100000028581/\thisrowno{1}-100000028581/\thisrowno{2})}] {amd_taylor_future_hpx_median.csv};

\legend{Arm \cpp{std::future}, Arm \cpp{hpx::future}, Intel \cpp{std::future}, Intel \cpp{hpx::future},AMD \cpp{std::future}, AMD \cpp{hpx::future}};
\end{axis}
\end{tikzpicture}
    }
      }
      \subfloat[\cpp{hpx::for_each}\label{fig:performance:amd}]{

      \resizebox{0.5\textwidth}{!}
        {
    \begin{tikzpicture}
\begin{axis}[grid,xlabel={\# cores},ylabel={Flop/s},xmax = 20,legend pos=north west]

\addplot[black,thick,mark=*, error bars/.cd, y dir=both, y explicit] table [x expr=\thisrowno{0},y expr={100000028581/\thisrowno{2}}, col sep=comma, y error expr={1*(100000028581/\thisrowno{1}-100000028581/\thisrowno{2})}] {ookami_taylor_par_hpx_median.csv};

\addplot[black,thick,mark=triangle*, error bars/.cd, y dir=both, y explicit] table [x expr=\thisrowno{0},y expr={100000028581/\thisrowno{2}}, col sep=comma, y error expr={1*(100000028581/\thisrowno{1}-100000028581/\thisrowno{2})}] {intel_taylor_par_hpx_median.csv};

\addplot[black,thick,mark=square*, error bars/.cd, y dir=both, y explicit] table [x expr=\thisrowno{0},y expr={100000028581/\thisrowno{2}}, col sep=comma, y error expr={1*(100000028581/\thisrowno{1}-100000028581/\thisrowno{2})}] {amd_taylor_par_hpx_median.csv};

\legend{Arm,Intel, AMD};
\end{axis}
\end{tikzpicture}
    }
}

      \subfloat[Arm 64FX \label{fig:performance:intel} ]{
  
            \resizebox{0.5\textwidth}{!}
        {
    \begin{tikzpicture}
\begin{axis}[grid,xlabel={\# cores},ylabel={Flop/s},xmax=20,legend pos=north west]

\addplot[black,thick,mark=*, error bars/.cd, y dir=both, y explicit] table [x expr=\thisrowno{0},y expr={100000028581/\thisrowno{2}}, col sep=comma, y error expr={1*(100000028581/\thisrowno{1}-100000028581/\thisrowno{2})}] {ookami_taylor_sender_receiver_hpx_median.csv};

\addplot[black,thick,mark=square*, error bars/.cd, y dir=both, y explicit] table [x expr=\thisrowno{0},y expr={100000028581/\thisrowno{2}}, col sep=comma, y error expr={1*(100000028581/\thisrowno{1}-100000028581/\thisrowno{2})}] {ookami_taylor_coroutine_hpx_median.csv};

\addplot[black,thick,mark=diamond*, error bars/.cd, y dir=both, y explicit] table [x expr=\thisrowno{0},y expr={100000028581/\thisrowno{2}}, col sep=comma, y error expr={1*(100000028581/\thisrowno{1}-100000028581/\thisrowno{2})}] {ookami_taylor_future_hpx_median.csv};

\addplot[black,thick,mark=triangle*, error bars/.cd, y dir=both, y explicit] table [x expr=\thisrowno{0},y expr={100000028581/\thisrowno{2}}, col sep=comma, y error expr={1*(100000028581/\thisrowno{1}-100000028581/\thisrowno{2})}] {ookami_taylor_par_hpx_median.csv};

\legend{Sender \& Receiver, Future + Coroutine, Future,\cpp{hpx:for_each}};
\end{axis}
\end{tikzpicture}
    }
      }
    \caption{The median out of ten runs with variations for various approaches.
    Futurization using \cpp{std::future} and \cpp{hpx::future}  \protect\subref{fig:performance:arm} and HPX's parallel algorithm using \cpp{hpx::for_each} \protect\subref{fig:performance:amd}. On Arm64FX coroutines and sender \& receiver were tested \protect\subref{fig:performance:intel}. To create an artificially work load, we computed the Taylor series in Equation~\eqref{eq:taylor:series} for $n=\num{1000000000}$ and measured \num{100000028581} floating point operations using \textit{perf} on a single Intel Core. Details on compilers and software versions are listed in Table~\ref{tab:software}.}
    \label{fig:performance}
\end{figure*}

\subsection{Additional HPX features}
For the \cpp{hpx::for_each} the performance in Figure~\ref{fig:performance:amd} on Intel and AMD is not a straight line, and we observe some rolling hills. Here, in this case, the default chunk size of one was used. Note that in the C\texttt{++} standard there is currently no option to specify the chunk size yet. HPX, however, does provide such an option, see Listing~\ref{algorithm:taylor:parallel:alg:hpx} in Section~\ref{sec:approaches:parallel:hpx:features}. Figure~\ref{fig:performance:chunk:size} shows the usage of the chunk sizes to make the scaling more linear. We use a dynamic chunk \cpp{hpx::execution::dynamic_chunck_size} size of \num{1e6}. Figure~\ref{fig:performance:chunk:size} shows the performance on Intel and AMD. For both architectures, the scaling behavior looks linear and the Flop\textbackslash s are a little bit higher. The additional features provided by HPX can affect the performance. However, these features are not yet in the C\texttt{++} standard.

\begin{figure}[tb]
    \centering

      \resizebox{0.5\textwidth}{!}
        {
    \begin{tikzpicture}
\begin{axis}[grid,xlabel={\# cores},ylabel={Flop/s},xmax = 20,legend pos=north west]

\addplot[black,thick,mark=*, error bars/.cd, y dir=both, y explicit] table [x expr=\thisrowno{0},y expr={100000028581/\thisrowno{2}}, col sep=comma, y error expr={1*(100000028581/\thisrowno{1}-100000028581/\thisrowno{2})}] {amd_taylor_par_hpx_median_dynamic.csv};

\addplot[black,thick,mark=triangle*, error bars/.cd, y dir=both, y explicit] table [x expr=\thisrowno{0},y expr={100000028581/\thisrowno{2}}, col sep=comma, y error expr={1*(100000028581/\thisrowno{1}-100000028581/\thisrowno{2})}] {intel_taylor_par_hpx_median_dynamic.csv};


\legend{AMD,Intel};
\end{axis}
\end{tikzpicture}
    }
    \caption{Running the parallel algorithms in Figure~\protect\ref{fig:performance:amd} using HPX's dynamic chunk size, see Section~\protect\ref{sec:approaches:parallel:hpx:features}, to have a linear scaling and some slightly better performance. However, this feature is not yet in the C\texttt{++} standard and is solely provided by HPX.}
    \label{fig:performance:chunk:size}
\end{figure}

\section{Conclusion}
\label{sec:conclusion}
We have shown that Modern C\texttt{++}, through its standard libraries and language features, provides a complete and expressive shared memory parallel programming infrastructure for a single node. Therefore, no external libraries or language extensions are necessary to write high-quality parallel C\texttt{++} applications. We sketched an example of how to use futures, coroutines, and parallel algorithms in the current C\texttt{++} standard based on a Taylor series code. Furthermore, we provided an introduction to senders and receivers, a framework that might be available in a future C\texttt{++} standard. For most of these programming mechanisms, we showcased the implementation using the C\texttt{++} Standard Library using system threads and using the C\texttt{++} library for concurrency and parallelism (HPX).
We did this because HPX provides a cutting-edge implementation of the parallel library proposals being considered by the C\texttt{++} standards committee.

A performance comparison on an Intel\textsuperscript{\textregistered} CPU, AMD CPU, and ARM\textsuperscript{\textregistered} A64FX demonstrates that the proposed parallel programming mechanisms do achieve portability of performance without code changes.

\section*{Supplementary materials}
The code for all examples is available on GitHub\textsuperscript{\textregistered}\footnote{\url{https://github.com/STEllAR-GROUP/parallelnumericalintegration}} or Zenodo\textsuperscript{\texttrademark}\footnote{\url{https://zenodo.org/record/7515618}}, respectively.

\section*{Compliance with Ethical Standards}

\paragraph{Funding}
The authors would like to thank Stony Brook Research Computing and Cyberinfrastructure, and the Institute for Advanced Computational Science at Stony Brook University for access to the innovative high-performance Ookami computing system, which was made possible by a \$5M National Science Foundation grant (\#1927880).

\paragraph{Disclosure of potential conflicts of interest}
The authors declare that they have no competing interests.

\paragraph{Research involving human participants and/or animals } This article does not contain any studies with human participants performed by any of the authors.

\paragraph{Informed consent}

Not applicable, since no humans were involved in our research.

\section{Acknowledgments}
We would also like to thank Alireza Kheirkhahan and the HPC admins who support the Deep Bayou cluster at Louisiana State University.


\bibliography{references.bib,new.bib}

\begin{thebibliography}{10}
\providecommand{\url}[1]{{#1}}
\providecommand{\urlprefix}{URL }
\providecommand{\doi}[1]{\url{https://doi.org/#1}}
\bibcommenthead

\bibitem{butenhof1997programming}
D.R. Butenhof, \emph{Programming with POSIX threads} (Addison-Wesley
  Professional, 1997)

\bibitem{cxx11_standard}
{C++ Standards Committee}, {ISO/IEC 14882:2011, Standard for Programming
  Language C++ (C++11)}.
\newblock Tech. rep., ISO/IEC JTC1/SC22/WG21 (the C++ Standards Committee)
  (2011).
\newblock \url{https://wg21.link/N3337}, last publicly available draft

\bibitem{cxx14_standard}
{C++ Standards Committee}, {ISO/IEC 14882:2014, Standard for Programming
  Language C++ (C++14)}.
\newblock Tech. rep., ISO/IEC JTC1/SC22/WG21 (the C++ Standards Committee)
  (2011).
\newblock \url{https://wg21.link/N4296}, last publicly available draft

\bibitem{cxx17_standard}
{The C++ Standards Committee}, {ISO International Standard ISO/IEC 14882:2017,
  Programming Language C++}.
\newblock Tech. rep., {Geneva, Switzerland: International Organization for
  Standardization (ISO).} (2017).
\newblock \urlprefix\url{http://www.open-std.org/jtc1/sc22/wg21}

\bibitem{chandra2001parallel}
R.~Chandra, L.~Dagum, D.~Kohr, R.~Menon, D.~Maydan, J.~McDonald, \emph{Parallel
  programming in OpenMP} (Morgan kaufmann, 2001)

\bibitem{Leiserson2011}
C.E. Leiserson, \emph{Cilk} (Springer US, Boston, MA, 2011), pp. 273--288.
\newblock \doi{10.1007/978-0-387-09766-4_289}.
\newblock \urlprefix\url{https://doi.org/10.1007/978-0-387-09766-4_289}

\bibitem{CarterEdwards20143202}
H.C. Edwards, et~al., {K}okkos: {E}nabling manycore performance portability
  through polymorphic memory access patterns.
\newblock {Journal of Parallel and Distributed Computing} \textbf{74}(12), 3202
  -- 3216 (2014)

\bibitem{kale1993charm++}
L.V. Kale, S.~Krishnan, in \emph{Proceedings of the eighth annual conference on
  Object-oriented programming systems, languages, and applications} (1993), pp.
  91--108

\bibitem{wu2023quantifying}
N.~Wu, I.~Gonidelis, S.~Liu, Z.~Fink, N.~Gupta, K.~Mohammadiporshokooh,
  P.~Diehl, H.~Kaiser, L.V. Kale, in \emph{Euro-Par 2022: Parallel Processing
  Workshops: Euro-Par 2022 International Workshops, Glasgow, UK, August 22--26,
  2022, Revised Selected Papers} (Springer, 2023), pp. 5--16

\bibitem{chamberlain2007parallel}
B.L. Chamberlain, et~al., Parallel programmability and the chapel language.
\newblock {The International Journal of High Performance Computing
  Applications} \textbf{21}(3) (2007)

\bibitem{ebcioglu2004x10}
K.~Ebcioglu, et~al., in \emph{Proceedings of the International Workshop on
  Language Runtimes, OOPSLA}, vol.~30 (Citeseer, 2004)

\bibitem{zheng2014upc++}
Y.~Zheng, et~al., in \emph{2014 IEEE 28th International Parallel and
  Distributed Processing Symposium} (IEEE, 2014), pp. 1105--1114

\bibitem{thoman2018taxonomy}
P.~Thoman, et~al., A taxonomy of task-based parallel programming technologies
  for high-performance computing.
\newblock {The Journal of Supercomputing} \textbf{74}(4) (2018)

\bibitem{Kaiser2020}
H.~Kaiser, et~al., {HPX - The C++ Standard Library for Parallelism and
  Concurrency}.
\newblock Journal of Open Source Software \textbf{5}(53), 2352 (2020)

\bibitem{p2300}
M.~Dominiak, et~al.
\newblock std::execution (2022).
\newblock \urlprefix\url{https://wg21.link/p2300}

\bibitem{grubel2015performance}
P.~Grubel, et~al., in \emph{{2015 IEEE International Conference on Cluster
  Computing}} (IEEE, 2015), pp. 682--689

\bibitem{shirzad2019scheduling}
S.~Shirzad, et~al., in \emph{{2019 IEEE/ACM Workshop on Machine Learning in
  High Performance Computing Environments (MLHPC)}} (IEEE, 2019), pp. 31--43

\bibitem{khatami2017hpx}
Z.~Khatami, et~al., in \emph{{Proceedings of the Third International Workshop
  on Extreme Scale Programming Models and Middleware}} (2017), pp. 1--8

\bibitem{9651210}
S.~Yadav, et~al., in \emph{2021 IEEE/ACM 6th International Workshop on Extreme
  Scale Programming Models and Middleware (ESPM2)} (2021), pp. 20--29

\bibitem{kretz2012vc}
M.~Kretz, V.~Lindenstruth, {Vc: A C++ library for explicit vectorization}.
\newblock {Software: Practice and Experience} \textbf{42}(11), 1409--1430
  (2012)

\bibitem{diehl2023benchmarking}
P.~Diehl, S.R. Brandt, M.~Morris, N.~Gupta, H.~Kaiser.
\newblock Benchmarking the parallel 1d heat equation solver in chapel, charm++,
  c++, hpx, go, julia, python, rust, swift, and java (2023)

\bibitem{diehl2023simulating}
P.~Diehl, G.~Dai{\ss}, K.~Huck, D.~Marcello, S.~Shiber, H.~Kaiser,
  D.~Pfl{\"u}ger, {Simulating Stellar Merger using HPX/Kokkos on A64FX on
  Supercomputer Fugaku}.
\newblock arXiv preprint arXiv:2304.11002  (2023)

\end{thebibliography}

\end{document}